\documentclass[twocolumn,twoside,slac_two]{revtex4}
\usepackage{amsmath}
\usepackage{amssymb}
\usepackage{graphicx}
\usepackage{fancyhdr}
\begin{document}

\title{Planck Scale Cosmology and Resummed Quantum Gravity}

%

\author{B.F.L. Ward}
\affiliation{Department of Physics, Baylor University, Waco, TX 76798, USA}

\begin{abstract}
We show that, by using amplitude-based resummation techniques for 
Feynman's formulation of Einstein's theory, we get quantum field theoretic 
'first principles' predictions for the 
UV fixed-point values of the dimensionless gravitational 
and cosmological constants. Connections to the phenomenological asymptotic 
safety analysis of Planck scale cosmology by Bonanno and Reuter are discussed.
\end{abstract}

\maketitle

\thispagestyle{fancy}


\section{Introduction}
Sometime ago, Weinberg~\cite{wein2} pointed-out that quantum gravity 
may be asymptotically safe in that the UV behavior of the 
theory corresponds to a UV-fixed point with a finite 
dimensional critical surface so that the S-matrix 
only depends on a finite number of dimensionless parameters. 
Recently, Bonanno and Reuter~\cite{reuter1,reuter2} have shown, using a 
realization developed by Reuter~\cite{reuter-laut} of the idea via 
Wilsonian field space exact renormalization group methods, 
that one arrives at a purely Planck scale quantum mechanical formulation 
the inflationary cosmological scenario of Guth and 
Linde~\cite{guth,linde} -- this is very attractive as it opens the 
possibility of a deeper understanding of that scenario without the 
need of the hitherto unseen inflaton scalar field. In what follows,
using the new
resummed theory~\cite{bw1,bw2,bw2a,bw2b,bw2c,bw2d,bw2e,bw2f,bw2g,bw2h} 
of quantum gravity, which is 
based on Feynman's original approach~\cite{rpf1,rpf2} to the subject,
we recover the properties as used in Refs.~\cite{reuter1,reuter2} 
for the UV fixed point of quantum gravity with the
added results that we get 'first principles'
predictions for the fixed point values of
the respective dimensionless gravitational and cosmological constants
in their analysis.\par
The discussion proceeds as follows. In the next section we review the 
formulation of Einstein's theory by Feynman, as it is not generally familiar.
In Section 3, we present the elements of the resummed version of Feynman's formulation, resummed quantum gravity. Section 4 presents the applications
to Planck scale cosmology as it is formulated by Bonanno and Reuter~\cite{reuter1,reuter2}. Section 5 contains our concluding remarks.\par
\section{Feynman's Formulation of Einstein's Theory}
In Feynman's approach~\cite{rpf1,rpf2} to quantum gravity, the starting point is that the metric of space-time undergoes quantum field theory fluctuations just like all point-particle fields: we write the metric of space-time as $g_{\mu\nu}(x)=\eta_{\mu\nu}+2\kappa h_{\mu\nu}(x)$ where $\eta_{\mu\nu}=\text{diag}(1,-1,-1,-1)$ is the flat Minkowski space background metric and $\kappa=\sqrt{8\pi G_N}$ so that $h_{\mu\nu}(x)$ is the quantum field of the graviton when $G_N$ is 
Newton's constant. For definiteness and reasons of pedagogy, we specialize
the complete theory here, which is 
\begin{equation}
{\cal L}(x) = \frac{1}{2\kappa^2}\sqrt{-g}\left( R -2\Lambda\right)
           + \sqrt{-g} L^{\cal G}_{SM}(x)
\label{lgwrld1a}
\end{equation} 
where $R$ is the curvature scalar, $g$ is the determinant of the metric
of space-time $g_{\mu\nu}$, $\Lambda$ is the cosmological
constant and $\L^{\cal G}_{SM}(x)$ is the diffeomorphism invariant
form of the SM Lagrangian obtained from the well-known SM Lagrangian
in Ref.~\cite{bar-pass} by standard differential-geometric methods~\cite{bw1}, to the case of a single scalar field, the Higgs field $\varphi(x)$, 
with a rest mass set
at $m=120$~GeV~\cite{lewwg,lewwga}, in interaction with the graviton so that the relevant Lagrangian is now that already considered by Feynman~\cite{rpf1,rpf2} when ignore the small cosmological constant~\cite{cosm1} (we will re-instate it shortly):
{\small
\begin{equation}
\begin{split}
{\cal L}(x) &= -\frac{\sqrt{-g}}{2\kappa^2} R
            + \frac{\sqrt{-g}}{2}\left(g^{\mu\nu}\partial_\mu\varphi\partial_\nu\varphi - m_o^2\varphi^2\right)\\
            &= \frac{1}{2}{\big\{} h^{\mu\nu,\lambda}\bar h_{\mu\nu,\lambda} - 2\eta^{\mu\mu'}\eta^{\lambda\lambda'}
\bar{h}_{\mu_\lambda,\lambda'}\eta^{\sigma\sigma'}\\
&\bar{h}_{\mu'\sigma,\sigma'}{\big\}}
          + \frac{1}{2}{\big\{}\varphi_{,\mu}\varphi^{,\mu}-m_o^2\varphi^2 {\big\}} \\
&-\kappa {h}^{\mu\nu}{\big[}\overline{\varphi_{,\mu}\varphi_{,\nu}}+\frac{1}{2}m_o^2\varphi^2\eta_{\mu\nu}{\big{]}}\\
            & \quad - \kappa^2 [ \frac{1}{2}h_{\lambda\rho}\bar{h}^{\rho\lambda}{\big{(}} \varphi_{,\mu}\varphi^{,\mu} - m_o^2\varphi^2 {\big{)}} \\
&- 2\eta_{\rho\rho'}h^{\mu\rho}\bar{h}^{\rho'\nu}\varphi_{,\mu}\varphi_{,\nu}] + \cdots \\
\end{split}  
\label{eq1}
\end{equation}}\noindent
where $\varphi_{,\mu}\equiv \partial_\mu\varphi$.
We define
$\bar y_{\mu\nu}\equiv \frac{1}{2}\left(y_{\mu\nu}+y_{\nu\mu}-\eta_{\mu\nu}{y_\rho}^\rho\right)$ for any tensor $y_{\mu\nu}$.
The Feynman rules for this theory were 
already worked-out by Feynman~\cite{rpf1,rpf2}.
where we use his gauge, $\partial^\mu \bar h_{\nu\mu}=0$.
\par
Concerning the non-zero value of $\Lambda$,~$\Lambda/\kappa^2\sim (0.0024\; \text{eV})^4$~\cite{cosm1}, we see that it is so small on the EW scale represented by the Higgs mass that its main effect
in our loop corrections will be to provide an IR regulator for the graviton infrared (IR) divergences. This subtle point should be understood as follows.
Our non-zero value of $\Lambda$ means that the true background metric is that
of de Sitter, not that of Minkowski. We study the theory using the Minkowski
background as an approximate representation of the actual de Sitter one, adding in the required corrections when we probe that regime of space-time where the correction is significant: this is in the far IR where the effective graviton IR regulator mass, already noted by Feynman~\cite{rpf2}, represents the effect of the de Sitter curvature in our loop calculus. Thus, we are not in violation of the no-go theorems in Refs.~\cite{vandam,zak}.\par
\begin{figure*}[t]
\centering
\includegraphics[width=135mm]{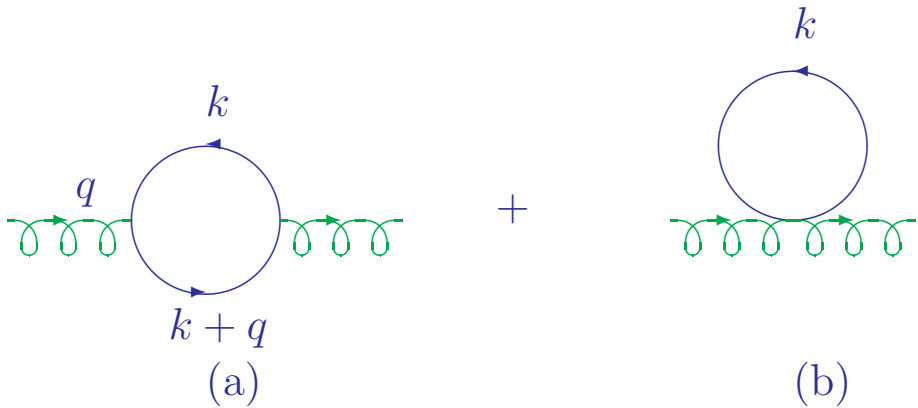}
\caption{The scalar one-loop contribution to the
graviton propagator. $q$ is the 4-momentum of the graviton.} \label{fig1}
\end{figure*}

The main stumbling block of the Feynman formulation is already evident in Fig.~\ref{fig1}, wherein we see that, by naive power counting, the graphs have superficial degree of divergence $D=4$, so that, even if we take gauge invariance into account, we still have $D_{eff}\ge 0$, and higher loops give higher values
of $D_{eff}$. The theory is thus, from this perspective, 
non-renormalizable as it is well-known.
\par
As we explain in Refs.~\cite{bw1,bw2,bw2a,bw2b,bw2c,bw2d,bw2e,bw2f,bw2g,bw2h}, 
this bad UV behavior can be
greatly improved by applying the methods of amplitude-based, 
exact resummation theory to arrive at what we have called resummed quantum gravity. We review this approach to the UV behavior of quantum gravity in the next section.\par 
\section{Resummed Quantum Gravity}
The basic strategy we use is to make an exact re-arrangement of the 
Feynman formulated perturbative series for Einstein's theory with the 
idea that the interactions
in the theory actually tame the attendant bad UV behavior 
dynamically. Intuitively, Newton's force is attractive between two positive masses, so that it becomes repulsive for negative mass-squared as we have in the deep Euclidean regime of the UV and this repulsion, in Feynman's overall space-time path-space approach, would lead to severe damping of UV propagation, thereby taming the otherwise bad UV behavior. This all would be consistent
with Weinberg's asymptotic safety approach as recently developed in Refs.~\cite{reuter1,reuter2,reuter-laut,reuter3,litim,perc}. As we have shown in Refs.~\cite{bw1}, exact resummation
of the IR dominated part of the proper self-energy function for a scalar
particle of mass $m$ gives the exact re-arrangement
\begin{equation}
i\Delta'_F(k)|_{\text{Resummed}} =  \frac{ie^{B''_g(k)}}{(k^2-m^2-\Sigma'_s+i\epsilon)}
\label{resum}
\end{equation}
where we have~\cite{bw1}
\begin{equation}
\begin{split} 
B''_g(k)&= -2i\kappa^2k^4\frac{\int d^4\ell}{16\pi^4}\frac{1}{\ell^2-\lambda^2+i\epsilon}\nonumber\\
&\qquad \frac{1}{(\ell^2+2\ell k+\Delta +i\epsilon)^2}
\end{split}
\label{yfs1} 
\end{equation}
when the use the IR regulator mass $\lambda$ for the graviton to represent
the leading effect of the small recently discovered~\cite{cosm1} 
cosmological constant, an effect Feynman already pointed-out 
in Ref.~\cite{rpf2}, for example. The residual self-energy function $\Sigma'_s$ starts in ${\cal O}(\kappa^2)$, so we may drop it in
calculating one-loop effects.\par
We note the following:\\
1. In the deep UV, explicit evaluation gives
\begin{equation}
B''_g(k) = \frac{\kappa^2|k^2|}{8\pi^2}\ln\left(\frac{m^2}{m^2+|k^2|}\right),
\label{deep}
\end{equation}
so that the resummed propagator falls faster than any power of $|k^2|$! 
Observe: 
in the Euclidean regime,
$-|k^2|=k^2$ so there is trivially no analyticity issue here.\\
2. If $m$ vanishes, using the usual $-\mu^2$ normalization point we get \;
$B''_g(k)=\frac{\kappa^2|k^2|}{8\pi^2}
\ln\left(\frac{\mu^2}{|k^2|}\right)$
which again vanishes faster than any power of $|k^2|$! 
This means that one-loop corrections are UV finite!
Indeed, as we show in Ref.~\cite{bw1}, 
all quantum gravity loops are UV finite!\\
3. In non-Abelian gauge theories,
the K\"all\'en-Lehmann representation cannot be used to show that 
the attendant gauge field renormalization constant $Z_3$
is formally less than 1 so that Weinberg's argument~\cite{wein2} that
the attendant spectral density condition, in an obvious notation, 
$\rho_{\text{K-L}}(\mu)\ge 0$ prevents the
graviton propagator from falling faster
than $1/k^2$ does not hold in such theories, as he has intimated
himself.\\
4. One might think that Ward-Takahashi identities would require that
the vertex correction resummation compensate any propagator resummation so that the net effect in a loop calculation if both vertices and propagators 
are resummed
is to leave the power counting in the UV for the loop unchanged~\cite{jpol1}.
In fact, if we put the square root of the propagator as a factor for each leg entering or leaving a vertex and resum as well the corresponding large IR effects in the vertex, we still have exponential damping because the large resummed IR effects in the vertex behave sub-dominantly~\cite{elswh} in the deep UV and this does not cancel the propagator fall-off.\\
5. The fact that we find that the dynamics of quantum gravity leads to UV finiteness is consistent with both the asymptotic safety approach of Weinberg,
as recently developed by Refs.~\cite{reuter1,reuter2,reuter-laut,reuter3,litim,perc} and with the recent 
leg renormalizable result of Kreimer~\cite{kreimer1}, wherein he finds
at least for the pure gravity part of Einstein's theory, using the Hopf-algebraic Dyson-Schwinger equation realization of renormalization theory~\cite{kreimer2}, that, while quantum gravity is non-renormalizable order by order in perturbation theory, there is an infinite set of relations among residues of the respective amplitudes so that when all are imposed only a finite number of unknown constants obtain, i.e., he finds in this way more evidence that quantum gravity is non-perturbatively renormalizable.\par
We have called our representation
of the quantum theory of general relativity resummed quantum gravity (RQG).
A number of applications have been worked-out in Refs.~\cite{bw1,bw2,bw2a,bw2b,bw2c,bw2d,bw2e,bw2f,bw2g,bw2h}.
We turn to its implications~\cite{bwi} for Planck scale cosmology in the next section.\par
\section{Planck Scale Cosmology}

\begin{figure*}[t]
\begin{center}
\includegraphics[width=135mm]{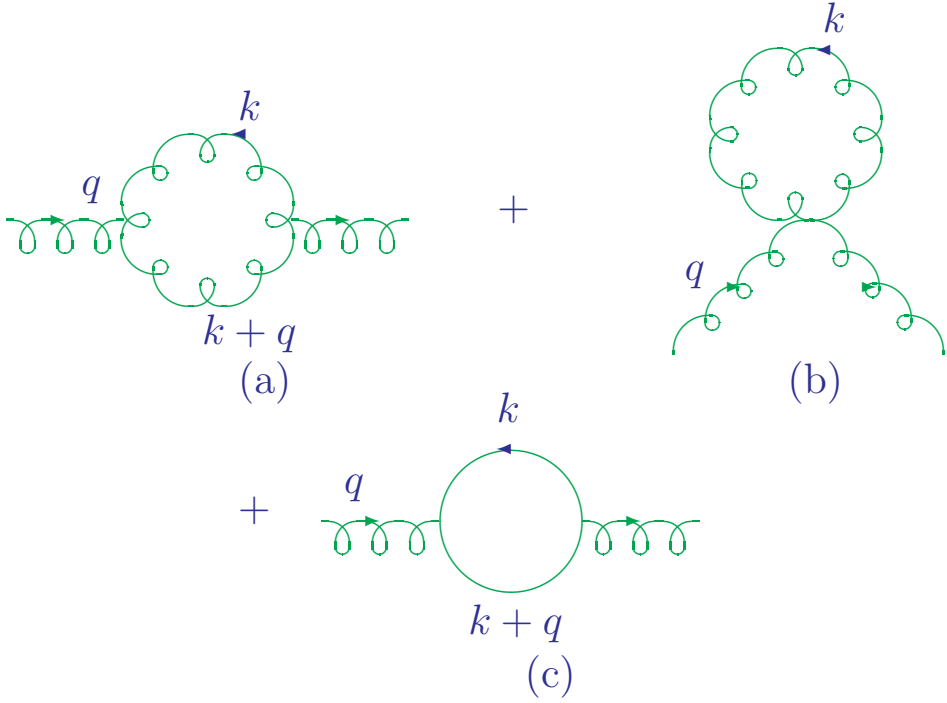}
\end{center}
\caption{The graviton((a),(b)) and its ghost((c)) one-loop contributions to the graviton propagator. $q$ is the 4-momentum of the graviton.}
\label{fig11}
\end{figure*}

Consider the graviton propagator in the theory of
gravity coupled to a massive scalar(Higgs) field~\cite{rpf1,rpf2}. We have the
graphs in Fig.~\ref{fig11} in addition to that in Fig.~\ref{fig1}.
Using the resummed theory, we get that the Newton potential
becomes
\begin{equation}
\Phi_{N}(r)= -\frac{G_NM}{r}(1-e^{-ar}),
\label{newtnrn}
\end{equation}
for
\begin{equation}
a \cong  0.210 M_{Pl},
\end{equation}
so that we have $$G(k)=G_N/(1+\frac{k^2}{a^2})\qquad \qquad $$,
which implies fixed point behavior for $k^2\rightarrow \infty$,
in agreement with the asymptotic safety approach of Weinberg as recently
developed in Refs.~\cite{reuter1,reuter2,reuter-laut,reuter3,litim,perc}. Indeed, in Refs.~\cite{bw1,bw2,bw2a,bw2b,bw2c,bw2d,bw2e,bw2f,bw2g,bw2h}, 
we have shown that we are in agreement with the results 
in Refs.~\cite{reuter1,reuter2,reuter-laut,reuter3,litim,perc} 
on several aspects of the UV 
limit of quantum gravity, such as the final state of Hawking 
radiation~\cite{hawk1,hawk2} for an originally very massive black hole.
Let us note for completeness that Ref.~\cite{bojo} gets a similar result in loop quantum gravity~\cite{lpqg}. 
Here we show that we also agree with the Planck scale cosmology
phenomenology developed in Refs.~\cite{reuter1,reuter2}.
We believe this strengthens 
the case for asymptotic safety.\par
Specifically, Bonanno and Reuter~\cite{reuter1,reuter2}
present a phenomenological approach to Planck scale cosmology wherein the
starting point is the Einstein-Hilbert theory
\begin{equation}
{\cal L}(x) = \frac{1}{2\kappa^2}\sqrt{-g}\left( R -2\Lambda\right).
\label{lgwrld1-1}
\end{equation} 
Using the phenomenological exact renormalization group
for the Wilsonian coarse grained effective 
average action in field space, the authors in 
Refs.~\cite{reuter1,reuter2,reuter3} show that
attendant running Newton constant $G_N(k)$ and running 
cosmological constant
$\Lambda(k)$ approach UV fixed points as $k$ goes to infinity
in the deep Euclidean regime  --
$k^2G_N(k)\rightarrow g_*,\; \Lambda(k)\rightarrow \lambda_*k^2$
for $k\rightarrow \infty$ in the Euclidean regime.
Due to the thinning of the degrees of freedom in
Wilsonian field space renormalization theory, the 
arguments of Ref.~\cite{foot} 
are obviated~\cite{bflwc}.\par
The contact with cosmology then proceeds as follows: 
invoking a phenomenological
 connection between the momentum scale $k$ characterizing the coarseness
 of the Wilsonian graininess of the average effective action and the
 cosmological time $t$, 
 the authors in Ref.~\cite{reuter1,reuter2} show the standard cosmological
 equations admit the following extension:
\begin{align}
(\frac{\dot{a}}{a})^2+\frac{K}{a^2}&=\frac{1}{3}\Lambda+\frac{8\pi}{3}G_N\rho\\
\dot{\rho}+3(1+\omega)\frac{\dot{a}}{a}\rho&=0\\
\dot{\Lambda}+8\pi\rho\dot{G_N}&=0\\
G_N(t)&=G_N(k(t))\\
\Lambda(t)&=\Lambda(k(t))
\label{coseqn1}
\end{align}
in a standard notation for the density $\rho$ and scale factor $a(t)$
with the Robertson-Walker metric representation as
\begin{equation}
ds^2=dt^2-a(t)^2\left(\frac{dr^2}{1-Kr^2}+r^2(d\theta^2+\sin^2\theta d\phi^2)\right)
\label{metric1}
\end{equation}
where $K=0,1,-1$ corresponds respectively  flat, spherical and
pseudo-spherical 3-spaces for constant time t for a
linear relation between the pressure $p$ and $\rho$
\begin{equation} 
p(t)=\omega \rho(t).
\end{equation}
The functional relationship between the respective
momentum scale $k$ and the cosmological 
time $t$ is determined
phenomenologically via
\begin{equation}
k(t)=\frac{\xi}{t}
\end{equation}
with the positive constant $\xi$ determined phenomenologically
.\par
Using the phenomenological, exact renormalization
group (asymptotic safety) UV fixed points as discussed above for $k^2G_N(k)=g_*$ and
$\Lambda(k)/k^2=\lambda_*$ 
the authors in Refs.~\cite{reuter1,reuter2}
show that the system in (\ref{coseqn1}) admits, for $K=0$,
a solution in the Planck regime ($0\le t\le t_{\text{class}}$, with
$t_{\text{class}}$ a few times the Planck time $t_{pl}$), which joins
smoothly onto a solution in the classical regime ($t>t_{\text{class}}$)
which agrees with standard Friedmann-Robertson-Walker phenomenology
but with the horizon, flatness, scale free Harrison-Zeldovich spectrum,
and entropy problems solved by Planck scale quantum physics.\par  
The fixed-point results $g_*,\lambda_*$ depend on the cut-offs
used in the Wilsonian coarse-graining procedure. 
The key properties of $g_*,\lambda_*$ used for the analysis in Refs.~\cite{reuter1,reuter2}(hereafter referred to as the B-R analysis) are that
they are both positive and that the product 
$g_*\lambda_*$ is cut-off/threshold function independent.
Here, we present the predictions for these
UV limits as implied by resummed quantum gravity theory, providing a more rigorous basis for the B-R analysis.\par 
Specifically, in addition to our UV fixed-point result for $G_N(k)\rightarrow a^2G_N/k^2\equiv g_*/k^2$,
we also get UV fixed point behavior for $\Lambda(k)$: using Einstein's equation
\begin{equation}
G_{\mu\nu}+\Lambda g_{\mu\nu}=-\kappa^2 T_{\mu\nu}
\end{equation}
and the point-splitting definition 
\begin{equation}
\begin{split}
\varphi(0)\varphi(0)&=\lim_{\epsilon\rightarrow 0}\varphi(\epsilon)\varphi(0)\cr
&=\lim_{\epsilon\rightarrow 0} T(\varphi(\epsilon)\varphi(0))\cr
&=\lim_{\epsilon\rightarrow 0}\{ :(\varphi(\epsilon)\varphi(0)): + <0|T(\varphi(\epsilon)\varphi(0))|0>\}\cr
\end{split}
\end{equation}
we get for a scalar the contribution to $\Lambda$, in Euclidean representation,
\begin{equation}
\begin{split}
\Lambda_s&=-8\pi G_N\frac{\int d^4k}{2(2\pi)^4}\frac{(2\vec{k}^2+2m^2)e^{-\lambda_c(k^2/(2m^2))\ln(k^2/m^2+1)}}{k^2+m^2}\cr
&\cong -8\pi G_N[\frac{3}{G_N^{2}64\rho^2}],\;\;\rho=\ln\frac{1}{\lambda_c}\cr
\end{split}
\end{equation} 
with $\lambda_c=\frac{2m^2}{M_{Pl}^2}$.
For a Dirac fermion, we get $-4$ times this contribution.\par
From these results, 
we get the Planck scale limit
\begin{equation}
\begin{split}
\Lambda(k) &\rightarrow k^2\lambda_*,\cr
\lambda_*&=\frac{1}{960\rho_{avg}}(\sum_jn_j)(\sum_{j}(-1)^{F_j}n_j)
\end{split}
\end{equation} 
where $F_j$ is the fermion number of $j$, $n_j$ is the effective
number of degrees of freedom of $j$,  and $\rho_{avg}$ is the average
value of $\rho$ -- see Ref.~\cite{bwi}.
\par
All of the Planck scale cosmology 
results of Bonanno and Reuter~\cite{reuter1,reuter2}
hold, but with definite results for the limits $k^2G(k)=g_*$ and
$\lambda_*$ for $k^2\rightarrow \infty$: solution of
the horizon and flatness problem,
scale free spectrum of primordial density fluctuations, initial entropy, etc.,
all provided by Planck scale quantum physics. \par
For reference, our UV fixed-point calculated here, 
$(g_*,\lambda_*)\cong (0.0442,0.232)$, can be compared with the estimates
of B-R, $(g_*,\lambda_*)\approx (0.27,0.36)$, with the understanding
that B-R analysis did not include SM matter action and that the
attendant results have definitely cut-off function
sensitivity. The qualitative results that $g_*$ and $\lambda_*$ are 
both positive and are significantly less than 1 in size with $\lambda_*>g_*$
are true of our results as well. We argue that this puts the 
results in Refs.~\cite{reuter1,reuter2} on a more firm theoretical basis.
\par
\section{Summary}
In this discussion, we have shown that the application of exact amplitude-based
resummation
methods, where we stress that for the 1PI 2-point function for example 
we have resummed the IR part of its loops
in Feynman's formulation of Einstein's theory for arbitrary values of the
respective external line momenta, we achieve the first {\it first principles} calculations of the UV limits of the dimensionless gravitational and cosmological constants. We have shown that these results agree with those found by the 
phenomenological asymptotic safety based exact, Wilsonian field space renormalization group analysis of Refs.~\cite{reuter1,reuter2,reuter-laut,reuter3,litim,perc} 
and that our results
support the properties of these limits as they are 
used in Refs.~\cite{reuter1,reuter2}
to formulate Planck scale cosmology as an alternative to the
standard inflationary cosmological paradigm of 
Guth and Linde~\cite{guth,linde}. We believe
our analysis puts the arguments in Refs.~\cite{reuter1,reuter2} for such an alternative on a more firm theoretical basis. 
Ultimately, we do expect experiment to make a choice between the two.\par 
\begin{acknowledgments}
We thank Profs. L. Alvarez-Gaume and W. Hollik for the support and kind
hospitality of the CERN TH Division and the Werner-Heisenberg-Institut, MPI, Munich, respectively, where a part of this work was done.
Work partly supported
by the US Department of Energy grant DE-FG02-05ER41399
and by NATO Grant PST.CLG.980342.
\end{acknowledgments}

\end{document}